\documentclass[aps,prd,twocolumn,showpacs,superscriptaddress,groupedaddress,nofootinbib]{revtex4}  
\usepackage{graphicx}  
\usepackage{dcolumn}   
\usepackage{xcolor}

\newcommand{\cR}{{\cal R}}

\usepackage{graphicx}
\usepackage{amsmath}
\usepackage{amsfonts}
\usepackage{epsfig}
\usepackage{slashed}
\usepackage{braket}

\usepackage{comment}
\usepackage{cases}

\def\bec{\begin{center}}
\def\eec{\end{center}}
\def\beq{\begin{equation}}
\def\eeq{\end{equation}}
\def\bea{\begin{eqnarray}}
\def\eea{\end{eqnarray}}
\def\KD{K\"{a}hler-Dirac }

\def\Psib{\overline{\Psi}}
\def\Phib{\overline{\Phi}}
\def\psib{\overline{\psi}}

\begin{document}
\title{Induced topological gravity and anomaly inflow from \KD fermions in odd dimensions}

\author{Simon Catterall}
\affiliation{Department of Physics, Syracuse University, Syracuse, NY 13244, USA }
\author{Arnab Pradhan}
\affiliation{Department of Physics, Syracuse University, Syracuse, NY 13244, USA }

\date{\today}

\begin{abstract}We show that the effective action that results from integrating out massive
\KD fermions propagating on a curved three dimensional space is a topological gravity theory of Chern-Simons
type. In the presence of
a domain wall, massless, two dimensional \KD fermions appear that are localized to the 
wall. Potential gravitational anomalies arising for these domain wall fermions
are cancelled via
anomaly inflow from the bulk gravitational theory. We also study
the invariance of the theory under large
gauge transformations.  The analysis and conclusions generalize
straightforwardly
to higher dimensions.
\end{abstract}

\pacs{}

\maketitle

\section{Introduction}
An alternative to the Dirac equation for describing fermions was proposed many years ago by K\"{a}hler \cite{kahler}. It is based on the simple observation that a natural square root of
the Laplacian is the operator $d-d^\dagger$ where $d$ is the exterior derivative and
$d^\dagger$ its adjoint. A key difference that distinguishes the \KD operator from its Dirac cousin
is the fact that the former can be defined without reference to a local frame and
spin connection. Thus \KD fermions are not globally equivalent to Dirac fermions and are
well defined on any smooth manifold.
Nevertheless there is a close connection between the two in flat space. The \KD equation
in $D$-dimensional flat space takes the form
\begin{equation}
    (d-d^\dagger-m)\Phi=0
\end{equation}
where $\Phi=\left(\omega,\omega_\mu,\omega_{\mu\nu},\ldots,\omega_{\mu_1\ldots\mu_D}\right)$ is a collection of $p$-form fields with $p$ running from $0$ to $D$. 
In this paper we will work in Euclidean space. It is straightforward to show that in even dimensions
this can be mapped into a Dirac equation describing $2^{D/2}$ degenerate
Dirac spinors corresponding to the columns of a $2^{D/2}\times 2^{D/2}$ matrix $\Psi$ \cite{Rabin:1981qj,Banks:1982iq}
\begin{equation}
    (\gamma^\mu\partial_\mu-m)\Psi=0
\end{equation}
where 
\begin{equation}
    \Psi=\sum_{\mu=0}^D \omega_{\mu_1\ldots\mu_p}\gamma^{\mu_1}\cdots\gamma^{\mu_p}.
    \label{expansion}\end{equation}
\KD fields arise naturally in twisted supersymmetric theories \cite{Catterall:2009it} and
are closely related to staggered fermions \cite{Becher:1982ud,Rabin:1981qj}. Recently, there has been renewed interest in them in connection 
with Dai-Freed anomalies \cite{Garcia-Etxebarria:2018ajm, Wan:2020ynf}, topological
insulators \cite{You:2014vea}
and symmetric mass generation in staggered fermion lattice models
\cite{Catterall:2015zua,Ayyar:2015lrd,Ayyar:2016lxq,Ayyar:2017qii,Catterall:2017ogi}.
They have also been proposed as an ingredient in the
construction of chiral lattice theories \cite{Catterall:2020fep}. 

One consequence of this
work has been the realization that
massless \KD theories in even dimensions suffer from a
gravitational anomaly,
which breaks a global $U(1)$
symmetry, unique to \KD fermions, down to $Z_4$ \cite{Catterall:2018lkj,Butt:2021brl}. In four dimensions this
anomaly is given by the Euler
density $\int \epsilon_{abcd} R^{ab}\wedge R^{cd}$ in contrast
to 
the usual gravitational anomaly of
Dirac fermions  given by
$\int R^{ab}\wedge R^{ab}$ with $R$ the Riemann tensor \cite{DELBOURGO1972381, PhysRevLett.37.1251}. It should be noted that since \KD fermions can be decomposed
into Dirac fermions in flat space they do not possess conventional $\gamma_5$ anomalies.

Remarkably this new
anomaly survives discretization since it depends only on the topology of the background which can be captured exactly in a simplicial approximation to the space. 
This $Z_4$ symmetry prohibits bare mass terms
but allows for four fermion interactions which can gap
fermions without breaking symmetries for multiples of two \KD fields \cite{Butt:2021brl}. In flat space each such \KD field
can be decomposed into $2^{D/2+1}$ 
Majorana spinors and we deduce that such theories contain
eight and sixteen Majorana spinors in two and four dimensions respectively. These magic fermion
numbers that allow for symmetric mass generation are in agreement with the 
cancellation of certain discrete anomalies for Weyl fermions - chiral fermion parity in two
dimensions and spin-$Z_4$ symmetry in four \cite{Garcia-Etxebarria:2018ajm,Razamat:2020kyf}. 

In this paper we will show that {\it massive} \KD fermions in odd dimensions exhibit further
interesting properties; they yield gravitational Chern-Simons theories at low energies. Furthermore, in the presence
of domain walls, these theories contain massless \KD fields localized
to the domain wall. We show that 
potential anomalies for these domain wall fermions,
of the type discussed above, are cancelled via anomaly inflow from the bulk
gravitational theory. 

\section{K\"{a}hler-Dirac fermions in three dimensions}
Following our earlier discussion the massless K\"{a}hler-Dirac (KD) action in three dimensions can be written as
\begin{equation}
    \int d^3x\;\sqrt{g}\; \Phib\left(d-d^\dagger\right)\Phi
\end{equation}
where $\Phi=(\phi,\phi_\mu,\phi_{\mu\nu},\phi_{\mu\nu\lambda})$ is a collection of $p$-forms
(antisymmetric tensors). Notice that such a field possesses
eight (complex) components in three dimensions. 

This action is invariant under a $U(1)$ symmetry of the form
\begin{align} \label{U1sym}
    \Phi&\to e^{i\alpha\Gamma}\Phi\\\nonumber
    \Phib&\to \Phib e^{i\alpha\Gamma}
\end{align}
where the linear operator $\Gamma$ acts on the component $p$-forms $\phi_p$ according to whether it carries an even or odd number of indices $\phi_p\to\left(-1\right)^p\phi_p$.
This property implies that $\Gamma$ anticommutes with the \KD operator which then ensures the $U(1)$ symmetry of the action. Furthermore, the operator $\Gamma$ can be used to construct projectors $P_\pm=\frac{1}{2}\left(I\pm\Gamma\right)$ which
act naturally on a \KD field to yield a pair of so-called reduced \KD fields
$\Phi_\pm=P_\pm\Phi$. The K\"{a}hler-Dirac operator maps between $\Phi_+$
to $\Phi_-$ and vice versa. 

If we want to map three dimensional \KD fermions into a set of spinors we will
need the analog of the matrix expansion given in eqn.~\ref{expansion}. Clearly one
cannot map the eight component fields of a \KD fermion in three dimensions using
just the minimal Dirac matrices corresponding to the
Pauli matrices. Instead one must double the number of components of
the spinor with the resulting matrix representation of
the three dimensional \KD field employing
$4\times 4$ gamma matrices~\footnote{Three dimensional fermions of this
type are called reducible fermions and correspond to a sum of the two irreducible spinor
representations for ${\rm spin}(3)$ - see \cite{Hands:2021mrg,Wipf:2022hqd}}. Naively such a field carries sixteen degrees of freedom but this can
be reduced to eight 
using the projection operators $P_\pm$ described earlier. These can be implemented in the matrix
representation as
\begin{equation}
    \Psi_\pm=P_\pm\Psi=\frac{1}{2}\left(\Psi\pm\gamma_5\Psi\gamma_5\right).
\end{equation}

The use of this four dimensional representation allows one
to write down a {\it massive} three dimensional \KD action which preserves the
$U(1)$ symmetry provided the mass term is taken proportional to $\gamma^4$:
\begin{equation}
    S=\int d^3x\,{\rm Tr}\left[\Psib\left(\gamma^\mu\partial_\mu-i\gamma^4M\right)P_+\Psi\right].
\end{equation}
This action is invariant under a global ${\rm spin}(3)$ Lorentz symmetry $L$ and
a global ${\rm spin}(4)$ flavor symmetry $F$ which act on the fields as 
\begin{align}
    \Psi_+&\to L\Psi_+ F^\dagger\nonumber\\
    \Psib_-&\to F\Psib_-L^\dagger.
\end{align}
Notice that $F$ should contain $L$ as a subgroup to reflect the \KD nature of the fermions since under a Lorentz transformation a \KD field must transform as
a sum of tensor representations. 
However, to facilitate the computation of the effective action in the next
section we will treat both symmetries as independent when gauging the action
and impose the \KD condition relating the corresponding gauge fields only after the
fermion integration.
On a curved space and having gauged the flavor symmetry the action is modified to \cite{Graf:1978kr}
\begin{equation}
    S=\int d^3x\, \hat{E}\,{\rm Tr}\left[\Psib(\hat{E}^\mu_\mathbb{A}\gamma^\mathbb{A} D_\mu-i\gamma^4M)P_+\Psi\right]\label{3dact}
\end{equation}
where $\hat{E}_\mu$ is a 3-frame corresponding to the background
metric and $D_\mu$ the associated covariant derivative which
acts on the field $\Psi$ as
\begin{equation}
D_\mu\Psi=\partial_\mu\Psi+\Omega_\mu\Psi-\Psi\hat{\Omega}_\mu\label{cov}
\end{equation}
where $\Omega_\mu$ is the three dimensional spin connection and $\hat{\Omega}_\mu$ 
is a ${\rm spin}(4)$ flavor gauge field. Notice that
while \KD fermions do not require the use of a spin connection it is necessary
to introduce such an object to do calculations in the matrix basis where the
\KD field is represented in terms of $\Psi$.

\section{Integrating out the fermions}\label{integrateout}
\begin{figure}[ht]
    \centering
    \includegraphics[width=.25\textwidth]{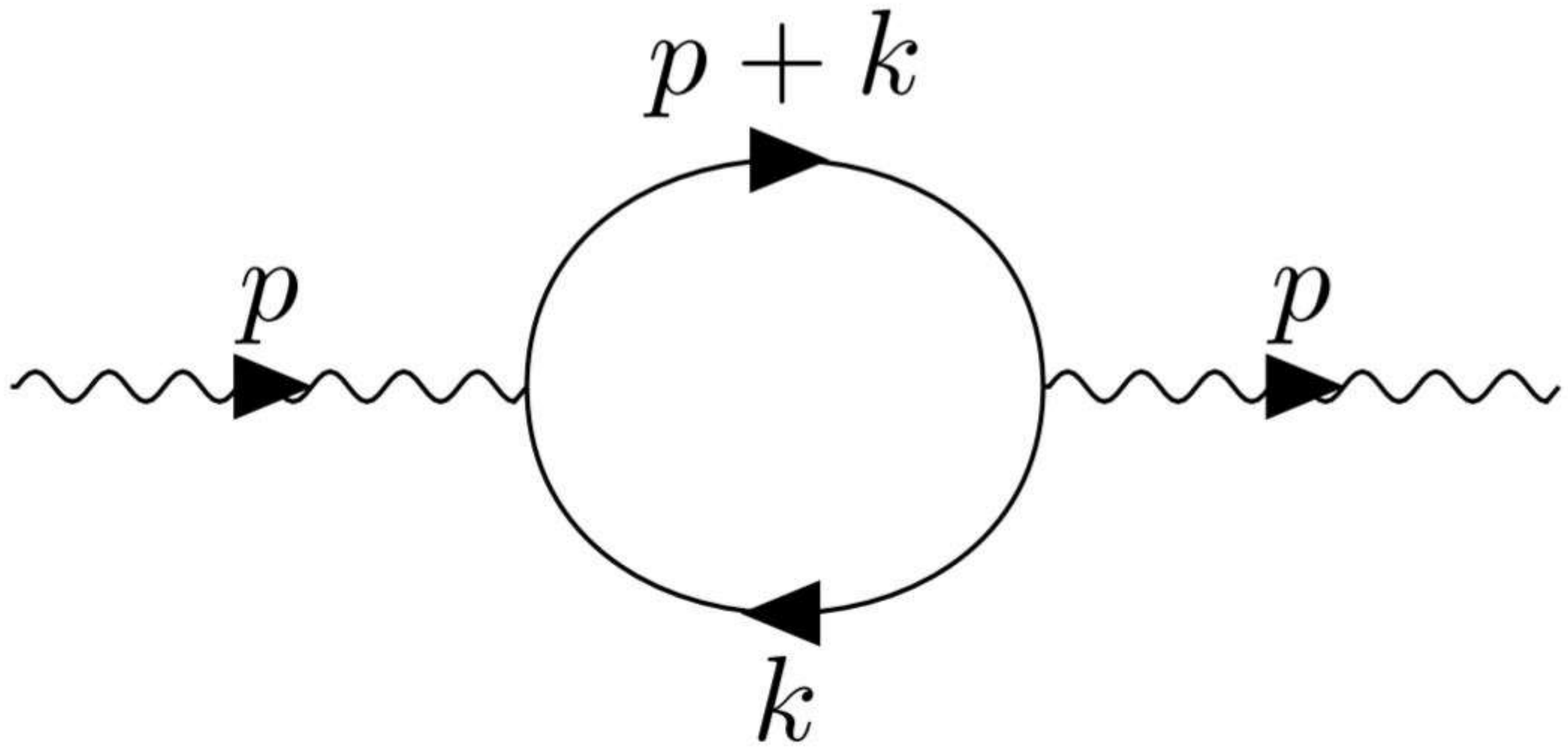}
    \caption{One loop contribution to vacuum polarization}
    \label{oneloop}
\end{figure}

\begin{figure}[ht]
    \centering
    \includegraphics[width=.25\textwidth]{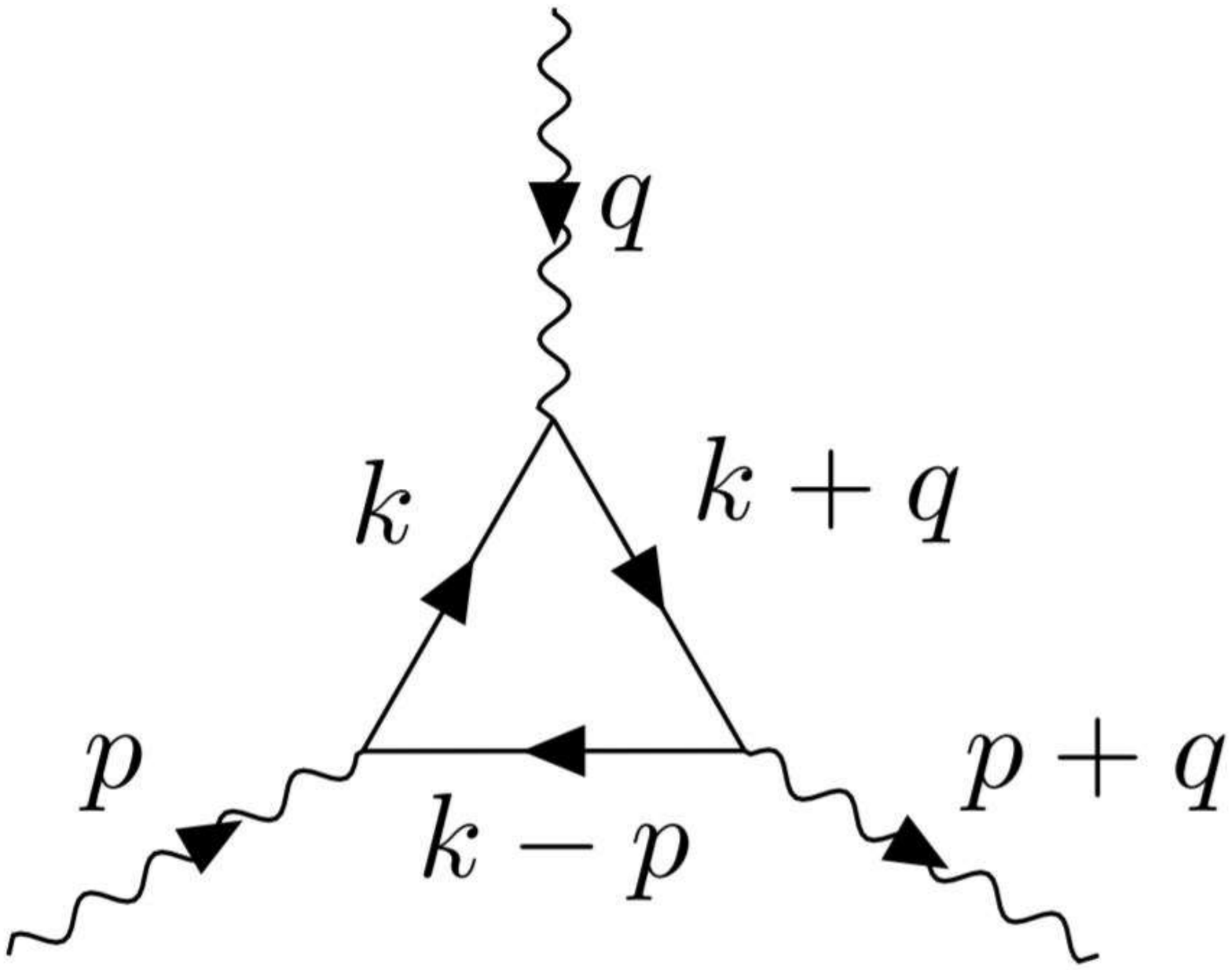}
    \caption{One loop contribution to three gauge boson vertex}
    \label{oneloop3}
\end{figure}

\noindent
We will focus on deriving an effective action 
for $\hat{\Omega}_\mu$ perturbatively in the limit $M\to\infty$. 
If we integrate out the fermions we obtain an effective action which can
be written
\begin{align}
    S_{eff}&= \text{Tr log}\left[(\slashed{\partial} -i \gamma^{4}M + \slashed{V})P_{+}\right]\\\nonumber
    &=\text{Tr }
    \text{log}\left[\left(\slashed{\partial} -i \gamma^{4}M\right)\left(I +
    \frac{\slashed{V}}{\slashed{\partial} -i \gamma^{4}M}
    \right)P_+\right]\\\nonumber
    \begin{split}
    &=\text{Tr log}\left[\left(I +
    \frac{\slashed{V}}{\slashed{\partial} -i \gamma^{4}M}
    \right)P_+\right]\\
    &\qquad +{\rm terms\;independent\;of\;\it V}.
    \end{split}
    \end{align}
    Expanding the logarithm  the leading term is 
    \begin{align}
    -\frac{1}{2}\text{Tr}\left[\left( \frac{\slashed{V}}{\slashed{\partial} -i \gamma^{4}M}\right)^2P_+\right]
\end{align}
corresponding to the diagram in fig~\ref{oneloop}.
Here $\slashed{V}=V_L+V_R$ where the subscripts 
indicate whether the gauge field acts on the left or right of the matrix
fermion $\Psi$:
\begin{align}
    V_L&=\hat{E}^\mu_{\mathbb{A}}\gamma^{\mathbb{A}}\Omega_\mu\\
    V_R&=-\hat{E}^\mu_{\mathbb{A}}\gamma^{\mathbb{A}}\hat{\Omega}_\mu.
\end{align}
In momentum space this gives~\footnote{For more details see appendix}
\begin{align}
\begin{split}
    S_{eff}^{quad} &= \frac{1}{2}\int \frac{d^3 k}{(2\pi)^3} \text{Tr}\bigg(\frac{\slashed{k}-\gamma^4 M}{k^2 + M^2}\gamma^\mu V_\mu  \frac{\slashed{k} + \slashed{p}-\gamma^4 M}{(k+p)^2 + M^2}\\
    &\qquad  \ \ \ \ \times\gamma^\nu V_\nu P_{+}\bigg).
\end{split}
\end{align}
If we focus on the
contribution that is linear in the external momentum $p$ we find
    \begin{align}\label{quadtrace}
    \begin{split}
     S_{eff}^{quad}   &= -\frac{1}{2}\times\frac{1}{2}\times\left(-\frac{1}{2}\right)^2 \text{tr}\left(\hat{E}\gamma^5\gamma^4 \gamma^{\mathbb{E}}\gamma^{\mathbb{F}}\gamma^{\mathbb{G}}\right)\\
     &\quad \times \hat{E}^\mu_{\mathbb{E}} \hat{E}^\delta_{\mathbb{F}} \hat{E}^\nu_{\mathbb{G}} \times \mathcal{I}\times p_{\delta}\hat{\Omega}_{\mu}^{\mathbb{A}\mathbb{B}}(-p)\hat{\Omega}_{\nu}^{\mathbb{C}\mathbb{D}}(p)\\
     &\qquad \times\text{tr}(\sigma_{\mathbb{A}\mathbb{B}}\sigma_{\mathbb{C}\mathbb{D}}\gamma_5).
     \end{split}
    \end{align}
Notice that 
a non-vanishing contribution comes only from employing $V_R$
at both vertices.  The integral $\mathcal{I}$ is given by
\begin{equation}
  \mathcal{I} = \int \frac{d^3 k}{(2\pi)^3} \frac{M}{(k^2 + M^2)((k+p)^2 + M^2)}.  
\end{equation}
For $M \gg p$ and rescaling $k/M \rightarrow k$
\begin{align}\label{piintegral}
  \mathcal{I} &= \frac{M}{|M|} \times \int^{\infty}_{-\infty} \frac{d^3 k}{(2\pi)^3} \frac{1}{(k^2 + 1)^2}\\\nonumber
  &= \frac{M}{|M|} \times \frac{1}{8\pi}.
  \end{align}
Employing the identity $\hat{E}\epsilon^{\mathbb{A}\mathbb{B}\mathbb{C}}\hat{E}^\mu_\mathbb{A}\hat{E}^\delta_\mathbb{B} \hat{E}^\nu_\mathbb{C} =\epsilon^{\mu\delta\nu}$
we find a contribution to the effective action of the form
\begin{align}\label{quad3d}
\begin{split}
    S_{ eff}^{quad} &= -i\frac{M}{|M|} \times 4 \times \left(\frac{1}{2}\right)^4 \times \frac{1}{8\pi} \\
    &\quad\times \int d^3x\,\epsilon^{\mu\delta\nu}\hat{\Omega}_{\mu}^{\mathbb{A}\mathbb{B}}(\partial_{\delta}\hat{\Omega}_{\nu}^{\mathbb{C}\mathbb{\mathbb{D}}})\epsilon_{\mathbb{A}\mathbb{B}\mathbb{C}\mathbb{D}}
    \end{split}
\end{align}
This is not gauge invariant. There is however a contribution coming from next order in the expansion of
the logarithm:
\begin{equation}
    \frac{1}{3}\text{Tr}\left[\left( \frac{\slashed{V}}{\slashed{\partial} -i \gamma^{4}M}\right)^3P_+\right]
\end{equation}
which corresponds to the Feynman diagram in fig.~\ref{oneloop3}.
In momentum space this gives
\begin{align}
\begin{split}
    S^{cubic}_{eff} &= \frac{i}{3}\int \frac{d^3 k}{(2\pi)^3} \text{Tr}\bigg(\frac{\slashed{k}-\gamma^4 M}{k^2 + M^2}\gamma^\mu V_\mu \frac{\slashed{k} + \slashed{q}-\gamma^4 M}{(k+q)^2 + M^2}\\
    &\qquad\times\gamma^\delta V_\delta \frac{\slashed{k} - \slashed{p} -\gamma^4 M}{(k-p)^2 +M^2}\gamma^\nu V_\nu P_{+}\bigg).
    \end{split}
    \end{align}
Extracting the leading term which again uses only $V_R$ yields 
\begin{align}
    \begin{split}
        S^{cubic}_{eff} &= -i\times\frac{1}{3}\times\frac{1}{2}\times\left(-\frac{1}{2}\right)^3 \text{tr}\left(\hat{E}\gamma^5\gamma^4 \gamma^{\mathbb{G}}\gamma^4\gamma^{\mathbb{H}}\gamma^4\gamma^{\mathbb{I}}\right)\\
        &\quad\times\hat{E}^\mu_\mathbb{G}\hat{E}^\delta_\mathbb{H} \hat{E}^\nu_\mathbb{I} \times \mathfrak{I}\times \hat{\Omega}_{\mu}^{\mathbb{A}\mathbb{B}}(-p)\hat{\Omega}_{\delta}^{\mathbb{C}\mathbb{D}}(-q)\hat{\Omega}_{\nu}^{\mathbb{E}\mathbb{F}}(p+q)\\
        &\qquad\times\text{tr}(\sigma_{\mathbb{A}\mathbb{B}}\sigma_{\mathbb{C}\mathbb{D}}\sigma_{\mathbb{E}\mathbb{F}}\gamma_5)\\
    &= i\times 4\times \frac{1}{3}\times\left(\frac{1}{2}\right)^4 \times \epsilon^{\mu\delta\nu} \times\mathfrak{I}\times \hat{\Omega}_{\mu}^{\mathbb{A}\mathbb{B}}(-p)\hat{\Omega}_{\delta}^{\mathbb{C}\mathbb{D}}(-q)\\
    &\quad\times\hat{\Omega}_{\nu}^{\mathbb{E}\mathbb{F}}(p+q) \times \frac{1}{8}(2 {\delta}_{\mathbb{A}\mathbb{B}} {\epsilon }_{\mathbb{C}\mathbb{D}\mathbb{E}\mathbb{F}}-3 {\delta}_{\mathbb{A}\mathbb{C}} {\epsilon }_{\mathbb{B}\mathbb{D}\mathbb{E}\mathbb{F}}\\
    &\qquad+{\delta}_{\mathbb{B}\mathbb{C}} {\epsilon }_{\mathbb{A}\mathbb{D}\mathbb{E}\mathbb{F}}+3 {\delta}_{\mathbb{A}\mathbb{D}} {\epsilon }_{\mathbb{B}\mathbb{C}\mathbb{E}\mathbb{F}}-{\delta}_{\mathbb{B}\mathbb{D}} {\epsilon }_{\mathbb{A}\mathbb{C}\mathbb{E}\mathbb{F}}\\
    &\qquad -2 {\delta}_{\mathbb{B}\mathbb{E}} {\epsilon }_{\mathbb{A}\mathbb{C}\mathbb{D}\mathbb{F}}-{\delta}_{\mathbb{C}\mathbb{E}} {\epsilon }_{\mathbb{A}\mathbb{B}\mathbb{D}\mathbb{F}}+{\delta}_{\mathbb{D}\mathbb{E}} {\epsilon }_{\mathbb{A}\mathbb{B}\mathbb{C}\mathbb{F}}\\
    &\qquad+2 {\delta}_{\mathbb{B}\mathbb{F}} {\epsilon }_{\mathbb{A}\mathbb{C}\mathbb{D}\mathbb{E}}+{\delta}_{\mathbb{C}\mathbb{F}} {\epsilon }_{\mathbb{A}\mathbb{B}\mathbb{D}\mathbb{E}}-{\delta}_{\mathbb{D}\mathbb{F}} {\epsilon }_{\mathbb{A}\mathbb{B}\mathbb{C}\mathbb{E}})\\
    &= i\times 4\times\frac{1}{3} \times\left(\frac{1}{2}\right)^4 \times \epsilon^{\mu\delta\nu}\times \mathfrak{I}\times \hat{\Omega}_{\mu}^{\mathbb{A}\mathbb{M}}(-p)\\
    &\quad\times\hat{\Omega}_{\delta}^{\mathbb{M}\mathbb{B}}(-q)\hat{\Omega}_{\nu}^{\mathbb{C}\mathbb{D}}(p+q)(2\epsilon_{\mathbb{A}\mathbb{B}\mathbb{C}\mathbb{D}})
    \end{split}
\end{align}
where
\begin{equation}
  \mathfrak{I} = \int \frac{d^3 k}{(2\pi)^3} \frac{M(k^2 + M^2)}{(k^2 + M^2)((k+q)^2 + M^2)((k-p)^2 +M^2)}.  
\end{equation}
For $M \gg p,q$ and rescaling $k/M \rightarrow k$
\begin{align}
  \mathfrak{I} &= \frac{M}{|M|} \times \int^{\infty}_{-\infty} \frac{d^3 k}{(2\pi)^3} \frac{1}{(k^2 + 1)^2} = \mathcal{I}
  \end{align}
where $\mathcal{I}$ is given by eqn.~\ref{piintegral}. In real space this yields
\begin{align}\label{cubic3d}
\begin{split}
    S_{ eff}^{cubic} &= i\times\frac{M}{|M|} \times 4\times \frac{1}{3}\times\left(\frac{1}{2}\right)^4 \times \frac{1}{8\pi}\\
    &\quad\times \int d^3x\,\epsilon^{\mu\delta\nu} \hat{\Omega}_{\mu}^{\mathbb{A}\mathbb{M}}\hat{\Omega}_{\delta}^{\mathbb{M}\mathbb{B}}\hat{\Omega}_{\nu}^{\mathbb{C}\mathbb{D}}(2\epsilon_{\mathbb{A}\mathbb{B}\mathbb{C}\mathbb{D}}).
\end{split}    
\end{align}
Combining eqn.~\ref{cubic3d} and eqn.~\ref{quad3d} gives the effective action
\begin{align}
\begin{split}
S^{CS}_{ eff}& =  -\frac{M}{|M|}\times \frac{i}{4 \times 8\pi} \int d^3x\,\epsilon^{\mu\delta\nu}\epsilon_{\mathbb{A}\mathbb{B}\mathbb{C}\mathbb{D}} \\
&\quad\times\left( \hat{\Omega}_{\mu}^{\mathbb{A}\mathbb{B}}\,(\partial_{\delta}\hat{\Omega}_{\nu}^{\mathbb{C}\mathbb{D}}) - \frac{2}{3} \hat{\Omega}_{\mu}^{\mathbb{A}\mathbb{M}}\hat{\Omega}_{\delta}^{\mathbb{M}\mathbb{B}}\hat{\Omega}_{\nu}^{\mathbb{C}\mathbb{D}}\right)\\
& =-\frac{M}{|M|}\times \frac{i}{32\pi}\int d^3x\, \epsilon^{\mu\delta\nu}\epsilon_{\mathbb{A}\mathbb{B}\mathbb{C}\mathbb{D}}\\
&\quad\times\hat{\Omega}_\mu^{\mathbb{A}\mathbb{B}}\left(\frac{F^{\mathbb{C}\mathbb{D}}_{\delta\nu}}{{2}}+\frac{1}{3}\hat{\Omega}^{\mathbb{C}\mathbb{M}}_\delta\hat{\Omega}^{\mathbb{M}\mathbb{D}}_\nu\right)\label{CSaction}
\end{split}
\end{align}
where $F$ is the ${\rm spin}(4)$ curvature. It is the
unique term in the effective action that survives the large $M$ limit.
Notice that while this piece of the
effective action comes from a U.V convergent integral 
this is not true of other terms arising in $S_{eff}^{CS}$ at finite $M$. Employing a Pauli-Villars
regulator with mass $\Lambda$ leads to the replacement 
$\frac{M}{|M|} \rightarrow \left(\frac{M}{|M|} + \frac{\Lambda}{|\Lambda|}\right)$
in eqn.~\ref{CSaction}. This modification
plays an important role in our later discussion
of domain wall physics and invariance of the effective action
under large gauge transfromations
in section VII.

We now impose the condition
that the original Lorentz symmetry be a subgroup of the ${\rm spin}(4)$ flavor
symmetry by setting
\begin{equation}
    \hat{\Omega}_\mu=\Omega_\mu^{AB}T_{AB} +2 E_\mu^{A}T_{4A}\quad A,B=1\ldots 3\label{top}
\end{equation}
where $\Omega_\mu$ is the original spin connection with $T_{AB} = \frac{1}{4}[\gamma_A,\gamma_B]$ the generators while $E_\mu$ are the additional gauge fields
needed for ${\rm spin}(4)$. In the next section we will see that $E_\mu$ can be interpreted as
a dynamical frame for an emergent geometry. Eq. \ref{CSaction} is hence a Chern-Simons term that ensures the effective action on a manifold
without boundary is invariant under gauge transformations of the
spin connection that can be smoothly deformed to the identity.

\section{Gravity interpretation}
We can decompose the ${\rm spin}(4)$ curvature also under the original 
Lorentz group by computing the commutator of the corresponding ${\rm spin}(4)$ covariant derivative $[D_\mu, D_\nu]$ and expanding the resultant
expression on the generators in a manner similar to that given in eqn.~\ref{top}.
This leads to the following expression:
\begin{equation}
    F_{\mu\nu}=\left(\cR_{\mu\nu}^{AB}-\frac{2}{\ell^2}E_{\left[\mu\right.}^A E_{\left.\nu\right]}^B\right)T_{AB}+
    \frac{4}{\ell}D_{\left[\mu\right.} E_{\left. \nu\right]}^A T_{4A}
\end{equation}
where $\cR_{\mu\nu}=\partial_\mu \Omega_\nu-\partial_\nu\Omega_\mu+[\Omega_\mu,\Omega_\nu]$ is the ${\rm spin}(3)$
curvature and the remaining components $D_{\left[\mu\right.} E_{\left. \nu\right]}$ are recognized as the torsion $T_{\mu\nu}$. Notice that
we have rescaled the gauge fields $E_\mu$ by an arbitrary length scale $\ell$ to make it possible to interpret $E_\mu$ as the dimensionless emergent frame.
Substituting these expressions into eqn.~\ref{CSaction} yields
\begin{equation}
\begin{split}
    S_{ eff}^{CS}&=-i\frac{M}{|M|}\times \frac{1}{32\pi}\int d^3x\,\epsilon^{\mu\nu\lambda}\epsilon_{ABC}\\
    &\quad\times\frac{1}{\ell}E_\mu^A\left(\cR_{\nu\lambda}^{BC}
    -\frac{8}{3\ell^2}E_\nu^B E_\lambda^C\right)
    \end{split}
\end{equation}
where we have discarded
boundary terms.
Clearly the action rewritten in these variables contains
both an Einstein-Hilbert and cosmological constant term as expected of a gravity
theory \cite{Chamseddine:1990gk,Zanelli:2005sa}. However, the 
relative coefficients of these terms have been fixed by the requirement
that the theory actually enjoys a local ${\rm spin}(4)$ 
symmetry now interpreted as a local de Sitter gauge symmetry.
Notice that the equation of motion for the Chern-Simons theory $F_{\mu\nu}=0$ now
implies the pair of equations
\begin{align}
    \cR_{\mu\nu}-\frac{2}{\ell^2}E_{\left[\mu\right.} E_{\left.\nu\right]}&=0,\\\nonumber
    T_{\mu\nu}&=0
\end{align}
corresponding to classical Euclidean de Sitter space and a torsion free connection.
Of course this
identification between Einstein Hilbert and Chern-Simons theory
is still problematic
at the non-perturbative level since in the path integral
the latter necessarily includes
degenerate metrics with vanishing frame. This is the origin of the topological
character of the gravity theory as discussed in \cite{Witten:2007kt}.

It is not a surprise that integrating out fermions in odd dimensions leads to
a Chern-Simons theory - this is well known in the case of
Dirac fermions transforming under some internal symmetry.
What is new here is that if those fermions are taken to be of \KD type propagating
on a curved background geometry then the induced Chern-Simons theory is actually
a (topological) theory of gravity.

\section{Domain wall construction}
\noindent
In the previous section we assumed that the three dimensional manifold was
compact. It is interesting to ask what happens in the presence of a boundary
or equivalently if a domain wall is introduced in the system. Our argument parallels
the original discussion by Callan and Harvey for Dirac fermions and later employed by
Kaplan in his construction of domain wall lattice fermions \cite{Callan:1984sa,Kaplan:1992bt}.

\begin{figure}[ht]
\centering
    \includegraphics[width=.35\textwidth]{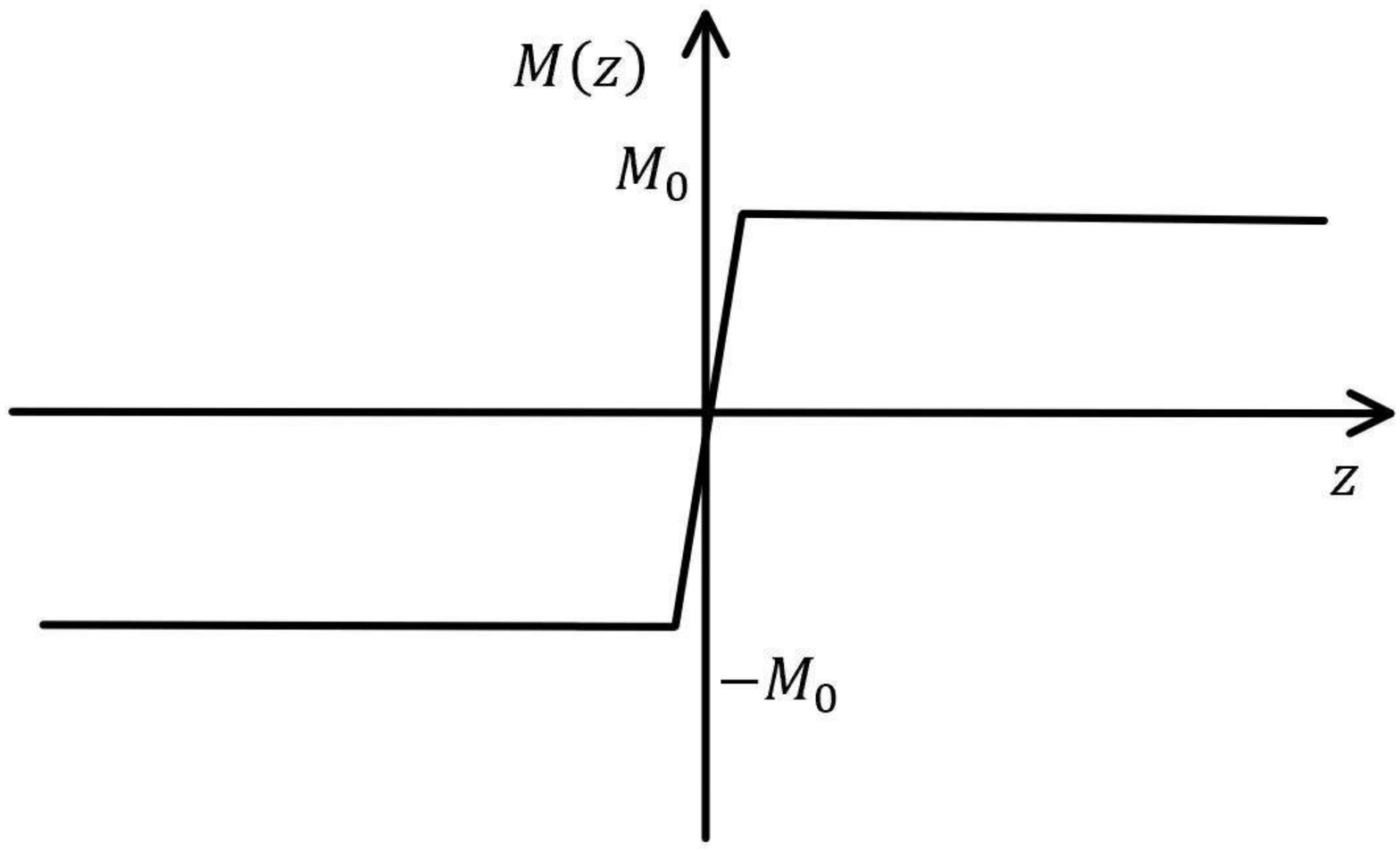}
    \caption{Domain wall}
    \label{3d_cubic}
\end{figure}

Let us imagine a manifold of the form ${\cal M}\times R$ with coordinates
$(x_\mu,z)$ where $x_\mu$ parameterize position on the
domain wall. Let us also allow the fermion mass $M$ to change sign as a function of 
the flat coordinate $z$ as shown in fig.~\ref{3d_cubic}
\begin{equation}
    M(z)=M_0 \frac{z}{|z|}.
\end{equation}
One expects that massless states appear at $z=0$. To see
this let us rewrite the bulk \KD equation in the form
\begin{equation}
    \left[\gamma^3\gamma^\mu D_\mu+\partial_z-i\gamma^3\gamma^4M(z)\right]\Psi(x,z)=0.
\end{equation}
One can find zero mode solutions of this equation
of the form
\begin{equation}
    \Psi_{\rm DW}=\chi(z)\psi(x)
\end{equation}
with $\gamma^3\gamma^\mu D_\mu\psi=0$
and $\psi(x)$ an eigenvector of the
hermitian operator $H=-i\gamma^3\gamma^4$ with eigenvalue $+1$.
The function $\chi(z)$ must then satisfy
\begin{equation}
    \partial_z\chi(z)=-M(z)\chi(z).
\end{equation}
Thus one finds $\chi(z)=e^{-M_0 |z|}$ corresponding to zero modes exponentially localized to the domain wall at $z=0$.
Notice that $\Psi_{\rm DW}$ contains just four degrees of freedom -- the original
reduced field $\Psi_+$ contained eight degrees of freedom while
the restriction to fields with $H=+1$ further halves the number of degrees of freedom. Four degrees of freedom corresponds to the field content of a two dimensional
\KD field propagating on the wall. We can verify this explicitly 
by going to a (Euclidean) chiral basis for the gamma matrices  corresponding
to 
\begin{equation}\gamma_\mu=\left(\begin{array}{cc}0&\sigma_\mu\\\overline{\sigma}_\mu&0\end{array}\right)\end{equation}
where $\sigma_\mu=(i\sigma_i,I)$ and $\overline{\sigma}_\mu=(-i\sigma_i,I)$. This implies that $\Psi_+$ takes the $2\times 2$ block form:
\begin{equation}
\Psi_+=\left(\begin{array}{cc}\psi_1&0\\0&\psi_2\end{array}\right).\label{psiblock}
\end{equation}
and the matrix $H$ takes the form:
\begin{equation}
 H=   \left(\begin{array}{cc}\sigma_3&0\\
    0&-\sigma_3\end{array}\right).\label{chiral}\end{equation}
The additional requirement that $\Psi_{\rm DW}$ be an eigenstate of $H$
with eigenvalue $+1$ shows that $\psi_1$ contains
two right handed two-dimensional Weyl spinors while $\psi_2$ contains two left handed spinors.

The constraint $H=1$ for the domain wall
fermions also breaks the original gauge symmetry to ${\rm spin}(2)\times {\rm spin}(2)$\footnote{Thus all gauge fields 
associated with the broken generators must vanish on
the domain wall.}. 
The first factor corresponds to the generator $\frac{1}{4}[\gamma^1,\gamma^2]$ and is associated
with the two dimensional spin connection $\hat{\Omega}_\mu^{12}$ needed to enforce local
Lorentz invariance for
the domain wall modes. The second factor corresponds 
to $H$ itself. The covariant derivative associated with rotations generated by $H$
takes the form
\begin{equation}
    D_\mu\Psi_{\rm DW}=\partial_\mu\Psi_{\rm DW} + \frac{i}{2}\hat{\Omega}^{34}_\mu\left(H\Psi_{\rm DW} -\Psi_{\rm DW}H\right).
\end{equation}
Using $H\Psi_{\rm DW}=\Psi_{\rm DW}$ this can be rewritten as
\begin{equation}
    D_\mu\Psi_{\rm DW}=\partial_\mu\Psi_{\rm DW} + \frac{i}{2}\hat{\Omega}^{34}_\mu\left(\Psi_{\rm DW} - H\Psi_{\rm DW}H\right).
\end{equation}
If we define the domain wall
chiral operator $\hat{\gamma}_5=\gamma_5 H$ the covariant derivative associated to $H$ becomes
\begin{equation}
    D_\mu\Psi_{\rm DW}=\partial_\mu \Psi_{\rm DW}^+ + \partial_\mu \Psi_{\rm DW}^- + i\hat{\Omega}^{34}_\mu\Psi_{\rm DW}^-
\end{equation}
where the domain wall reduced field $\Psi_{\rm DW}^\pm$ is given by
\begin{equation}
\Psi_{\rm DW}^\pm=\hat{P}_\pm\Psi_{\rm DW}=\frac{1}{2}\left(\Psi_{\rm DW}\pm \hat{\gamma}^5\Psi_{\rm DW}\hat{\gamma}^5\right). 
\end{equation}
Thus we find that the gauge field $\hat{\Omega}_{34}$ couples only to a two dimensional
reduced \KD field on the wall.

A similar feature is seen in the interaction of the domain wall fermion
with the two dimensional spin connection $\hat{\Omega}_{12}$. The corresponding term in
the covariant derivative takes the form:
\begin{align}
     &\frac{i}{2}\hat{\Omega}^{12}_\mu\left(i\gamma^1\gamma^2\Psi_{\rm DW} -\Psi_{\rm DW}i\gamma^1\gamma^2\right)\\\nonumber
     &=i\hat{\Omega}^{12}_\mu\hat{\gamma}^5\Psi_{\rm DW}^-.
\end{align}
Thus all gauge interactions on the domain wall couple only to the reduced \KD field $\Psi_{\rm DW}^-$. 

To summarise we find that the low energy excitations of the three dimensional
\KD theory in the
presence of a domain wall are massless
two dimensional \KD fermions $\Psi_{\rm DW}$ localized to the wall and described by a
Lorentz invariant action possessing an additional $U(1)$ symmetry generated
by an operator $\hat{\Gamma}=\hat{\gamma}^5\otimes \hat{\gamma}^5$.
The operator $\hat{\Gamma}$ anticommutes with the 
two dimensional \KD operator 
describing the domain wall fermions and allows the
\KD field to be projected into two independent reduced \KD fields $\Psi_{\rm DW}^-$ and
$\Psi_{\rm DW}^+$. Only one of these components $\Psi_{\rm DW}^-$ participates
in the remaining ${\rm spin}(2)\times {\rm spin}(2)$ gauge symmetry.

\section{Anomaly inflow for \KD fermions}
At first glance the structure of the domain wall fermion action
appears problematic since it is known that massless \KD fields in even dimensions suffer
from a gravitational anomaly \cite{Catterall:2018lkj,Butt:2021brl} that breaks
this $U(1)_{\hat{\Gamma}}$ symmetry down to $Z_4$. The two dimensional
domain wall action we derived
in the previous section includes a gauged
version of this symmetry (since the gauge field couples to a $\hat{\Gamma}$ reduced fermion) and 
hence one naively expects an anomaly induced breaking of gauge
invariance in the low energy theory. 
In this section we will show that there is
an additional contribution which arises from the bulk action which restores gauge invariance via an anomaly inflow mechanism.

To show in detail how this occurs we first
include a brief review of the derivation of the anomaly specialized
to the case of two dimensional domain wall fermions.
Under a $U(1)_{\hat{\Gamma}}$ rotation with parameter $\alpha(x)$ the
measure for the reduced \KD field $\Psi_{\rm DW}^-$ transforms by a factor $e^{i\int d^2x\,\alpha(x)A(x)}$ with
\begin{equation}
A(x)=\lim_{M\to\infty}{\rm Tr}\,\sum_n e\left(\overline{\phi}_n e^{\frac{1}{M^2}(\slashed{D})^2}\hat{P}_{-}P_{+}\phi_n\right)
\end{equation}
where we have regulated the UV divergence by inserting the factor $e^{\frac{1}{M^2}(\slashed{D})^2}$ where $\phi_n$ are eigenstates of the domain
wall \KD operator $\slashed{D}=\gamma^3\gamma^\mu D_\mu$ and $e$ represents the determinant of the frame restricted to
the wall which we denote as $e_\mu^a$.  Cyclically permuting the trace we find
\begin{align}
A&=\lim_{M\to\infty}{\rm Tr}\, \left( e^{\frac{1}{M^2}(\slashed{D})^2}\hat{P}_{-}P_{+}\sum_n e \phi_n \overline{\phi}_n\right)\\\nonumber
 &=\lim_{x\to x^{'}}\lim_{M\to\infty}{\rm Tr}\, \left( e^{\frac{1}{M^2}(\slashed{D})^2}\hat{P}_{-}P_{+}\delta(x - x^{'})\right).\end{align}
Expanding $\slashed{D}^2$ we obtain
\begin{align}
\begin{split}
    A&=\lim_{x\to x^{'}}\lim_{M\to\infty}-\frac{1}{4}\times{\rm Tr}\,\Big((-i\gamma^3\gamma^4)e\gamma^5\\
    &\quad\times e^{\frac{1}{M^2}(\Box + {\frac{1}{2}}e^\mu_a e^\nu_b {\sigma^{ab}}F_{\mu\nu}^{cd}[\sigma_{cd},.])}\\
    &\qquad \times \delta(x - x^{'})\gamma^5(-i\gamma^3\gamma^4)\Big)\\
    &= \lim_{x\to x^{'}}\lim_{M\to\infty}{\rm Tr}\,\Big((\sigma^{34})e\gamma^5\\
    &\quad\times e^{\frac{1}{M^2}(\Box + {\frac{1}{2}}e^\mu_a e^\nu_b {\sigma^{ab}}F_{\mu\nu}^{cd}[\sigma_{cd},.])}\\
    &\qquad \times \delta(x - x^{'})\gamma^5(\sigma_{34})\Big)
    \end{split}
\end{align}
where $F$ contains the surviving non-zero components of the
${\rm spin}(4)$ curvature corresponding to the symmetry ${\rm spin}(2)\times {\rm spin}(2)$.
Expanding the exponential to $\mathcal{O}(1/M^2)$ to get a non-zero result for the trace over spinor
and flavor indices  and acting with $e^{\frac{1}{M^2}\Box^2}$ on the delta
function yields
\begin{align}
\begin{split}
    A &=-\frac{1}{4\pi}\times\left(\frac{1}{2}\right) {\rm tr}\,
\left(e\gamma^5\sigma^{ab}\sigma^{34}\right)e^\mu_ae^\nu_b F_{\mu\nu}^{cd}\,{\rm tr}\,\left(\gamma_5\sigma_{cd}\sigma_{34}\right)\\
&= -\frac{1}{8\pi}\epsilon^{\mu\nu}\epsilon_{cd}
R_{\mu\nu}^{cd}
\label{anomaly}
\end{split}
\end{align}
where $R_{\mu\nu}$ corresponds to the curvature of the spin connection $\hat{\Omega}_{12}$. We have employed the result $e\epsilon^{ab}e^\mu_a e^\nu_b =\epsilon^{\mu\nu}$ in the last line. 
Hence, under a $U(1)$ transformation the 
measure for a reduced \KD field transforms as~\footnote{Taking $\alpha(x)$ to be a constant one finds the measure transforms
by the phase $e^{-i\chi\alpha}$ where $\chi$ is the Euler characteristic
of the two dimensional space. If one further replaces the reduced
field by a full \KD field and compactifies the space to $S^2$ where $\chi=2$
one obtains the original
$U(1)$ global anomaly referred to in the introduction. In such a background
the phase is just $e^{-4i\alpha}$ which leaves an unbroken $Z_4$ subgroup.}
\begin{equation}
    \int D\Psib\,D\Psi \rightarrow e^{-\frac{i}{8\pi}\int d^{2}x \,\alpha(x)\,\epsilon^{\mu\nu}\epsilon_{cd}R_{\mu\nu}^{cd}}\,\int D\Psib\,D\Psi.\label{chiralanom}
\end{equation}
This naively breaks gauge invariance.
However, the anomaly we have computed for the domain wall fermions is not the whole story.
We showed earlier that the bulk contains also an induced
Chern-Simons term. In general this also undergoes a non-zero change under a gauge
transformation. In general
the variation of the bulk Chern-Simons action takes the form:
\begin{align}
\begin{split}
    \delta S^{CS}_{ eff} &= -\frac{M}{|M|} \times \frac{i}{32\pi} \times \epsilon^{\mu\delta\nu}\epsilon_{\mathbb{A}\mathbb{B}\mathbb{C}\mathbb{D}}\\
    &\quad\times\left(\partial_{\mu}\hat{\Omega}_{\delta}^{\mathbb{A}\mathbb{B}} - \partial_{\delta}\hat{\Omega}_{\mu}^{\mathbb{A}\mathbb{B}} - 2\hat{\Omega}_{\mu}^{\mathbb{A}\mathbb{M}}\hat{\Omega}_{\delta}^{\mathbb{M}\mathbb{B}}\right) \delta\hat{\Omega}_{\nu}^{\mathbb{C}\mathbb{D}}\\
    &= -\frac{M}{|M|} \times \frac{i}{32\pi} \times \epsilon^{\mu\delta\nu}\epsilon_{\mathbb{A}\mathbb{B}\mathbb{C}\mathbb{D}}F^{\mathbb{A}\mathbb{B}}_{\mu\delta}\delta\hat{\Omega}_{\nu}^{\mathbb{C}\mathbb{D}}
\end{split}
\end{align}
where $F$ is the ${\rm spin}(4)$ curvature.
Under a gauge transformation $\hat{\Omega}_\mu^{\mathbb{A}\mathbb{B}} \rightarrow \hat{\Omega}_\mu^{\mathbb{A}\mathbb{B}}+D_\mu\zeta^{\mathbb{A}\mathbb{B}}$ 
the effective action changes:
\begin{align}\label{csdomain}
    \delta S^{CS}_{eff} &= -\int d^{3}x \frac{M}{|M|} \times \frac{i}{32\pi} \times \epsilon^{\mu\delta\nu}\epsilon_{\mathbb{A}\mathbb{B}\mathbb{C}\mathbb{D}}F^{\mathbb{A}\mathbb{B}}_{\mu\delta} (D_\nu\zeta^{\mathbb{C}\mathbb{D}})\nonumber\\
    &= i\int d^{3}x \frac{1}{16\pi} \times \epsilon^{\mu\delta}\epsilon_{\mathbb{A}\mathbb{B}\mathbb{C}\mathbb{D}}F^{\mathbb{A}\mathbb{B}}_{\mu\delta}\zeta^{\mathbb{C}\mathbb{D}}\partial_{z}\left(\frac{M}{2|M|}\right)\\\nonumber
    &= i\int d^{3}x \ \delta(z) \times \frac{1}{16\pi} \times \epsilon^{\mu\delta}\epsilon_{ab}\epsilon_{cd}F^{ab}_{\mu\delta}\zeta^{cd}
    \end{align}
where $a,b=\{1,2\}$ while $c,d=\{3,4\}$ and these indices
are to be contracted
using two independent two-dimensional $\epsilon$ symbols
corresponding to the product of the two invariant tensors for 
${\rm spin}(2)\times {\rm spin}(2)$ - the surviving symmetry on the domain wall. Taking $\zeta^{34}=-\zeta^{43}=\alpha(x)$ we find
\begin{equation}\label{csvari}
       \delta S^{CS}_{eff}= \frac{i}{8\pi}\int d^{2}x\, \alpha(x)\,\epsilon^{\mu\delta}\epsilon_{ab}R^{ab}_{\mu\delta}.
\end{equation}
Thus the gauge transformation of the Chern-Simons term in the presence of the domain
wall generates a contribution that is equal in magnitude but opposite in sign to
that coming from the anomalous variation of the measure for the domain wall fermions - eqn. \ref{chiralanom}. Thus the 
bulk and boundary variations cancel and the full theory is gauge invariant.
This is anomaly inflow
in action for \KD fields.  That this should occur is guaranteed by the fact
that the Euler characteristic of the bulk theory is zero if it is
taken to be a product of a two dimensional space and a circle since
$\chi(S^1)=0$~\footnote{Our previous discussion assumed z extends from $-\infty$ to $\infty$ but we can replace this by a circle at a price of adding an anti-domain wall at infinity.}. 

One might worry that the previous argument ignores the fact that the Chern-Simons
term was computed for constant mass which is certainly not the situation close to
the domain wall. However, it is possible to avoid this problem if one
simply computes the change in the Chern-Simons current 
$J^{34}_\mu=\frac{\delta S_{eff}}{\delta A_\mu^{34}}$ between $z=\infty$
and $z=-\infty$.
One then finds
\begin{equation}
 \Delta J^{34}_3= 2\times 2\times \frac{1}{32}\epsilon^{\mu\nu 3}\epsilon_{AB}R^{AB}_{\mu\nu}
\end{equation}
where the second factor of two arises from the double counting
associated with the fact that $A_\mu^{AB}=-A_\mu^{BA}$.
Comparing this to the divergence of the $U(1)$ current arising from the domain wall
fermions $\partial^\mu J_\mu^{34}=-\frac{1}{8\pi}\epsilon^{\mu\nu}\epsilon_{AB} R_{\mu\nu}^{AB}$
we see that the net flow of charge off the domain wall is accounted for by
the Chern-Simons current.

\section{Invariance under large gauge transformations}
It is of course interesting to ask about the 
invariance of the theory under large gauge transformations.
To facilitate this analysis it is convenient to again adopt a Euclidean chiral basis 
for the gamma matrices. The ${\rm spin}(4)$ connection becomes
\begin{equation}
\hat{\Omega}=
\Omega^i\left(\begin{array}{cc}i\sigma_i&0\\0&i\sigma_i\end{array}\right)+ E^i\left(\begin{array}{cc}i\sigma_i&0\\0&-i\sigma_i\end{array}\right)
\end{equation}
while the fermion field takes the form
\begin{equation}
    \Psi_+=\left(\begin{array}{cc}\psi_1&0\\
    0&\psi_2\end{array}\right)\qquad \Psib_-=\left(\begin{array}{cc}0&\psib_1\\
    \psib_2&0\end{array}\right).
\end{equation}
The \KD action then separates into two independent contributions
\begin{equation}
\begin{split}
    S&=\int d^3x\,\hat{E}\big[
    {\rm tr}\,\left(\psib_1\left(\slashed{D}(\Omega+E)+iM\right)\psi_1\right)\\
    &\quad +
    {\rm tr}\,\left(\psib_2\left(\slashed{D}(\Omega-E)-iM\right)\psi_2\right)\big]
    \end{split}
\end{equation}
where ${\rm tr}$ denotes a trace over a two dimensional block.
Each such block will then
generate its own $SU(2)$ Chern-Simons term on integration over the
fermions :
\begin{equation}
    S=I_{\rm CS}(\Omega+E)-I_{\rm CS}(\Omega-E)
    \label{CS3d}
\end{equation}
where
\begin{equation}
\begin{split}
 I_{\rm CS}(A)&=\frac{1}{32\pi}{\rm sign}(M)\int d^3x\,
 \epsilon^{\mu\nu\rho}\\
 &\quad \times {\rm tr}\left( A_\mu\partial_\nu A\rho-
 \frac{2}{3}A_\mu A_\nu A_\rho\right).
 \end{split}
\end{equation} 
The relative minus sign in eqn.~\ref{CS3d} 
arises because of differing signs of the mass in the two blocks \cite{Witten:2007kt}.

Under a gauge transformation 
$A_\mu \rightarrow g(x) A_\mu g^{-1}(x) + g(x) \partial_\mu g^{-1}(x)$, 
each Chern-Simons term transforms, up to a boundary term, according to
\begin{equation}
\begin{split}
\delta I_{CS}&=\int d^3 x \  \epsilon^{\mu\nu\delta} \left(\frac{1}{96\pi}\frac{M}{|M|}\right) {\rm tr}\left(g\partial_\mu g^{-1} g\partial_\nu g^{-1} g\partial_\delta g^{-1}\right)\\
&=\frac{M}{|M|} \pi n\label{lgt}
\end{split}
\end{equation}
where the winding number $n = \pi_3 (SU(2)) = \mathbb{Z}$\footnote{The normalization of our CS term
reflects the non-standard trace 
${\rm tr}(\sigma^a\sigma^b)=2\delta^{ab}$.}.

Thus naively the level number of the CS term is $k=\pm 1/2$ with
the partition function changing sign for odd $n$. However,
once one regulates the theory
with a Pauli-Villars field corresponding to a $z$-independent cut-off mass $\Lambda$ the coefficients of the Chern Simons
terms (the level numbers) are shifted to $k=0$ and $k=1$
in the two regions $z<0$ and $z>0$ respectively.  Thus we find that
the theory is in fact also
invariant under large gauge transformations.

\section{Summary}

We have shown that integrating out massive \KD 
fermions in a curved three dimensional background
yields a Chern-Simons term. This Chern-Simons term corresponds to a topological theory
of gravity in which both spin connection and frame emerge from an extended 
gauge symmetry - in this case Euclidean de Sitter symmetry. Gravity theories of this
type were proposed many years ago \cite{Chamseddine:1989nu,Chamseddine:1990gk,Zanelli:2005sa} 
and generalize Witten's old observation that three dimensional gravity can be formally written as a Chern-Simons
gauge theory \cite{Witten:1988hc}. 

In the presence of a
domain wall we have shown that massless two dimensional \KD fermions
appear on the wall. These are described by a single \KD field which
can be decomposed into two independent
components called reduced \KD fields which carry half the number of
degrees of freedom.  We find that just
one of these reduced fields participates in the
gauge interactions on the domain wall.
Furthermore although the reduced \KD fermions on the wall
suffer from a gravitational 
anomaly there is no violation
of local gauge invariance 
because of anomaly inflow from the bulk gravitational Chern-Simons
term.

It is not hard to generalize this construction to higher dimensions. For example,
the effective long distance action for massive \KD fermions in five dimensions is also a topological gravity theory of Chern-Simons type \cite{Chamseddine:1989nu,Chamseddine:1990gk,Zanelli:2005sa} with
gauge group ${\rm spin}(6)$ in Euclidean space.
Using the same arguments as for three dimensions it is 
clear that massless four
dimensional \KD fermions invariant under local ${\rm spin}(4)$ Lorentz transformations and
an additional local $U(1)$ symmetry would then
arise in the presence of a domain wall 
in such a theory. Again, the domain wall
action will contain a coupling of the $U(1)$ gauge field to a single reduced \KD field.
As in three dimensions 
gauge invariance of the theory remains intact since the gauge
variation of the five dimensional Chern-Simons term cancels the potentially
anomalous variation arising from the four dimensional reduced fermions.

One of the conclusions one can draw from our work is that coupling
reduced \KD fermions to gravity in some even dimensional space 
is inconsistent due to a (mixed) gravitational anomaly {\it unless}
the theory lives on a domain wall or 
boundary of a space of one higher dimension. If this additional dimension
is finite there will necessarily be an anti-domain wall which localizes another
reduced \KD fermion with the opposite eigenvalue of $\Gamma$. In this scenario
the Chern-Simons current naturally flows between the two walls and the low energy
theory is manifestly well defined.

Much of our discussion for \KD fermions has paralleled existing arguments for Dirac
fermions. In this paper we have focused on perturbative anomalies and anomaly
inflow. But in \cite{Butt:2021brl} it was shown that \KD fermions also
exhibit discrete 't Hooft anomalies. Cancelling these anomalies is a necessary
condition for symmetric mass generation and requires multiples of two
\KD fields or four reduced \KD fields. If we take the flat space limit this constraint translates
into the requirement that the theory contain sixteen Majorana spinors in
four dimensions in perfect
agreement with results for gapping edge modes in four dimensional topological
insulators which require cancellation of a seemingly unrelated
't Hooft anomaly for a spin-$Z_4$
symmetry acting on Weyl fermions. This makes it plausible that theories
of Weyl fermions, which are free of all 't Hooft anomalies
and hence capable of symmetric mass generation, can be written
in terms of \KD fermions. Furthermore since the anomalies of
\KD fermions survive intact under discretization this suggests that 
they may be important for constructing lattice mirror models that
target chiral theories in the continuum limit.
Indeed numerical simulations of two flavors of
interacting staggered fermions (which are obtained by
discretization of \KD fermions) show evidence for the existence
of a massive symmetric phase \cite{Butt:2018nkn}. Further work is needed
to understand these issues in more detail.

\acknowledgments
This work was supported by the US Department of Energy (DOE), Office of Science, Office of High Energy Physics under Award Number {DE-SC0009998}. The authors would like to thank A.P. Balachandran for comments on an early draft of the paper.

\appendix
\section{Details on the computation of $S_{eff}$}
     For a complete set of basis states $\phi_n$
    \begin{align}
    \begin{split}
    S_{eff}&= \int d^3 x \text{Tr}\sum_n \hat{E} \left(\overline{\phi}_n\text{log}[(\slashed{D} -i \gamma^{4}M )P_{+}]\phi_n\right)\\
    &= \int d^3 x \text{Tr} \left(\text{log}[(\slashed{D} -i \gamma^{4}M )P_{+}]\sum_n \hat{E} \phi_n\overline{\phi}_n\right)\\
    &= \lim_{x\to x'}\int d^3 x \text{Tr} \left(\text{log}[(\slashed{D} -i \gamma^{4}M )P_{+}]\delta(x-x')\right)\\
    &= \lim_{x\to x'}\int d^3 x \text{Tr} \bigg(\text{log}[(\slashed{D} -i \gamma^{4}M )P_{+}]\\
    &\quad\times\int\frac{d^3 k}{(2\pi)^3} e^{ik_\mu D^\mu \sigma(x,x')}\bigg)
    \end{split}
    \end{align}
where $\sigma(x,x')$ is the geodesic biscalar [a generalization of $\frac{1}{2}(x - x')^2$ in flat space] defined by
\begin{equation}
\sigma(x,x') = \frac{1}{2}g^{\mu\nu}D_\mu \sigma (x,x') D_\nu \sigma (x,x').
\end{equation}
Expanding the logarithm and exponential as power series and using the properties of the geodesic biscalar \cite{10.1143/PTP.81.512}
\[\sigma (x,x) = 0,\]
\[\lim_{x\to x'} D_\mu D_\nu \sigma (x,x') = g_{\mu\nu},\]

\[\text{and} \lim_{x\to x'} D_\mu D_\nu D_\alpha \sigma (x,x') = 0,\]
we get
\begin{align}
    S_{eff}= \int \hat{E} \ d^3 x\int \frac{1}{\hat{E}}\frac{d^3 k}{(2\pi)^3} \text{Tr} \left(\text{log}[(\slashed{k} -i \gamma^{4}M )P_{+}]\right)
\end{align}
where $\slashed{k} = g_{\mu\nu}\gamma^\mu k^\nu$. The determinant $\hat{E}$ has been restored to make the invariance of real-space and $k$-space measures manifest. We can now choose locally flat coordinates to evaluate the $k$-space integral. This
allows us to reduce the calculation of $S_{eff}$ to an equivalent flat space
problem:
\begin{equation}
\begin{split}
S_{eff} &\equiv \lim_{x\to x'}\int d^3 x \text{Tr} \bigg(\text{log}[(\slashed{\partial} -i \gamma^{4}M )P_{+}]\\
&\quad\times\int \frac{d^3 k}{(2\pi)^3} e^{ik_\mu (x^\mu - x'^{\mu})}\bigg).
\end{split}
\end{equation}

\bibliography{chiralKD}

\begin{thebibliography}{31}
\expandafter\ifx\csname natexlab\endcsname\relax\def\natexlab#1{#1}\fi
\expandafter\ifx\csname bibnamefont\endcsname\relax
  \def\bibnamefont#1{#1}\fi
\expandafter\ifx\csname bibfnamefont\endcsname\relax
  \def\bibfnamefont#1{#1}\fi
\expandafter\ifx\csname citenamefont\endcsname\relax
  \def\citenamefont#1{#1}\fi
\expandafter\ifx\csname url\endcsname\relax
  \def\url#1{\texttt{#1}}\fi
\expandafter\ifx\csname urlprefix\endcsname\relax\def\urlprefix{URL }\fi
\providecommand{\bibinfo}[2]{#2}
\providecommand{\eprint}[2][]{\url{#2}}

\bibitem[{\citenamefont{Kahler}((1962))}]{kahler}
\bibinfo{author}{\bibfnamefont{E.}~\bibnamefont{Kahler}}, \bibinfo{journal}{{
  Rend. Math 3-4}} \textbf{\bibinfo{volume}{{21}}}, \bibinfo{pages}{425}
  (\bibinfo{year}{(1962)}).

\bibitem[{\citenamefont{Rabin}(1982)}]{Rabin:1981qj}
\bibinfo{author}{\bibfnamefont{J.~M.} \bibnamefont{Rabin}},
  \bibinfo{journal}{Nucl. Phys. B} \textbf{\bibinfo{volume}{201}},
  \bibinfo{pages}{315} (\bibinfo{year}{1982}).

\bibitem[{\citenamefont{Banks et~al.}(1982)\citenamefont{Banks, Dothan, and
  Horn}}]{Banks:1982iq}
\bibinfo{author}{\bibfnamefont{T.}~\bibnamefont{Banks}},
  \bibinfo{author}{\bibfnamefont{Y.}~\bibnamefont{Dothan}}, \bibnamefont{and}
  \bibinfo{author}{\bibfnamefont{D.}~\bibnamefont{Horn}},
  \bibinfo{journal}{Phys. Lett.} \textbf{\bibinfo{volume}{B117}},
  \bibinfo{pages}{413} (\bibinfo{year}{1982}).

\bibitem[{\citenamefont{Catterall et~al.}(2009)\citenamefont{Catterall, Kaplan,
  and Unsal}}]{Catterall:2009it}
\bibinfo{author}{\bibfnamefont{S.}~\bibnamefont{Catterall}},
  \bibinfo{author}{\bibfnamefont{D.~B.} \bibnamefont{Kaplan}},
  \bibnamefont{and} \bibinfo{author}{\bibfnamefont{M.}~\bibnamefont{Unsal}},
  \bibinfo{journal}{Phys. Rept.} \textbf{\bibinfo{volume}{484}},
  \bibinfo{pages}{71} (\bibinfo{year}{2009}), \eprint{0903.4881}.

\bibitem[{\citenamefont{Becher and Joos}(1982)}]{Becher:1982ud}
\bibinfo{author}{\bibfnamefont{P.}~\bibnamefont{Becher}} \bibnamefont{and}
  \bibinfo{author}{\bibfnamefont{H.}~\bibnamefont{Joos}}, \bibinfo{journal}{Z.
  Phys. C} \textbf{\bibinfo{volume}{15}}, \bibinfo{pages}{343}
  (\bibinfo{year}{1982}).

\bibitem[{\citenamefont{Garc\'\i{}a-Etxebarria and
  Montero}(2019)}]{Garcia-Etxebarria:2018ajm}
\bibinfo{author}{\bibfnamefont{I.~n.} \bibnamefont{Garc\'\i{}a-Etxebarria}}
  \bibnamefont{and} \bibinfo{author}{\bibfnamefont{M.}~\bibnamefont{Montero}},
  \bibinfo{journal}{JHEP} \textbf{\bibinfo{volume}{08}}, \bibinfo{pages}{003}
  (\bibinfo{year}{2019}), \eprint{1808.00009}.

\bibitem[{\citenamefont{Wan and Wang}(2020)}]{Wan:2020ynf}
\bibinfo{author}{\bibfnamefont{Z.}~\bibnamefont{Wan}} \bibnamefont{and}
  \bibinfo{author}{\bibfnamefont{J.}~\bibnamefont{Wang}},
  \bibinfo{journal}{JHEP} \textbf{\bibinfo{volume}{07}}, \bibinfo{pages}{062}
  (\bibinfo{year}{2020}), \eprint{1910.14668}.

\bibitem[{\citenamefont{You and Xu}(2015)}]{You:2014vea}
\bibinfo{author}{\bibfnamefont{Y.-Z.} \bibnamefont{You}} \bibnamefont{and}
  \bibinfo{author}{\bibfnamefont{C.}~\bibnamefont{Xu}}, \bibinfo{journal}{Phys.
  Rev.} \textbf{\bibinfo{volume}{B91}}, \bibinfo{pages}{125147}
  (\bibinfo{year}{2015}), \eprint{1412.4784}.

\bibitem[{\citenamefont{Catterall}(2016)}]{Catterall:2015zua}
\bibinfo{author}{\bibfnamefont{S.}~\bibnamefont{Catterall}},
  \bibinfo{journal}{JHEP} \textbf{\bibinfo{volume}{01}}, \bibinfo{pages}{121}
  (\bibinfo{year}{2016}), \eprint{1510.04153}.

\bibitem[{\citenamefont{Ayyar and
  Chandrasekharan}(2016{\natexlab{a}})}]{Ayyar:2015lrd}
\bibinfo{author}{\bibfnamefont{V.}~\bibnamefont{Ayyar}} \bibnamefont{and}
  \bibinfo{author}{\bibfnamefont{S.}~\bibnamefont{Chandrasekharan}},
  \bibinfo{journal}{Phys. Rev. D} \textbf{\bibinfo{volume}{93}},
  \bibinfo{pages}{081701} (\bibinfo{year}{2016}{\natexlab{a}}),
  \eprint{1511.09071}.

\bibitem[{\citenamefont{Ayyar and
  Chandrasekharan}(2016{\natexlab{b}})}]{Ayyar:2016lxq}
\bibinfo{author}{\bibfnamefont{V.}~\bibnamefont{Ayyar}} \bibnamefont{and}
  \bibinfo{author}{\bibfnamefont{S.}~\bibnamefont{Chandrasekharan}},
  \bibinfo{journal}{JHEP} \textbf{\bibinfo{volume}{10}}, \bibinfo{pages}{058}
  (\bibinfo{year}{2016}{\natexlab{b}}), \eprint{1606.06312}.

\bibitem[{\citenamefont{Ayyar and Chandrasekharan}(2017)}]{Ayyar:2017qii}
\bibinfo{author}{\bibfnamefont{V.}~\bibnamefont{Ayyar}} \bibnamefont{and}
  \bibinfo{author}{\bibfnamefont{S.}~\bibnamefont{Chandrasekharan}},
  \bibinfo{journal}{Phys. Rev. D} \textbf{\bibinfo{volume}{96}},
  \bibinfo{pages}{114506} (\bibinfo{year}{2017}), \eprint{1709.06048}.

\bibitem[{\citenamefont{Catterall and Butt}(2018)}]{Catterall:2017ogi}
\bibinfo{author}{\bibfnamefont{S.}~\bibnamefont{Catterall}} \bibnamefont{and}
  \bibinfo{author}{\bibfnamefont{N.}~\bibnamefont{Butt}},
  \bibinfo{journal}{Phys. Rev. D} \textbf{\bibinfo{volume}{97}},
  \bibinfo{pages}{094502} (\bibinfo{year}{2018}), \eprint{1708.06715}.

\bibitem[{\citenamefont{Catterall}(2021)}]{Catterall:2020fep}
\bibinfo{author}{\bibfnamefont{S.}~\bibnamefont{Catterall}},
  \bibinfo{journal}{Phys. Rev. D} \textbf{\bibinfo{volume}{104}},
  \bibinfo{pages}{014503} (\bibinfo{year}{2021}), \eprint{2010.02290}.

\bibitem[{\citenamefont{Catterall et~al.}(2018)\citenamefont{Catterall, Laiho,
  and Unmuth-Yockey}}]{Catterall:2018lkj}
\bibinfo{author}{\bibfnamefont{S.}~\bibnamefont{Catterall}},
  \bibinfo{author}{\bibfnamefont{J.}~\bibnamefont{Laiho}}, \bibnamefont{and}
  \bibinfo{author}{\bibfnamefont{J.}~\bibnamefont{Unmuth-Yockey}},
  \bibinfo{journal}{JHEP} \textbf{\bibinfo{volume}{10}}, \bibinfo{pages}{013}
  (\bibinfo{year}{2018}), \eprint{1806.07845}.

\bibitem[{\citenamefont{Butt et~al.}(2021)\citenamefont{Butt, Catterall,
  Pradhan, and Toga}}]{Butt:2021brl}
\bibinfo{author}{\bibfnamefont{N.}~\bibnamefont{Butt}},
  \bibinfo{author}{\bibfnamefont{S.}~\bibnamefont{Catterall}},
  \bibinfo{author}{\bibfnamefont{A.}~\bibnamefont{Pradhan}}, \bibnamefont{and}
  \bibinfo{author}{\bibfnamefont{G.~C.} \bibnamefont{Toga}},
  \bibinfo{journal}{Phys. Rev. D} \textbf{\bibinfo{volume}{104}},
  \bibinfo{pages}{094504} (\bibinfo{year}{2021}), \eprint{2101.01026}.

\bibitem[{\citenamefont{Delbourgo and Salam}(1972)}]{DELBOURGO1972381}
\bibinfo{author}{\bibfnamefont{R.}~\bibnamefont{Delbourgo}} \bibnamefont{and}
  \bibinfo{author}{\bibfnamefont{A.}~\bibnamefont{Salam}},
  \bibinfo{journal}{Physics Letters B} \textbf{\bibinfo{volume}{40}},
  \bibinfo{pages}{381} (\bibinfo{year}{1972}), ISSN \bibinfo{issn}{0370-2693},
  \urlprefix\url{https://www.sciencedirect.com/science/article/pii/0370269372908258}.

\bibitem[{\citenamefont{Eguchi and Freund}(1976)}]{PhysRevLett.37.1251}
\bibinfo{author}{\bibfnamefont{T.}~\bibnamefont{Eguchi}} \bibnamefont{and}
  \bibinfo{author}{\bibfnamefont{P.~G.~O.} \bibnamefont{Freund}},
  \bibinfo{journal}{Phys. Rev. Lett.} \textbf{\bibinfo{volume}{37}},
  \bibinfo{pages}{1251} (\bibinfo{year}{1976}),
  \urlprefix\url{https://link.aps.org/doi/10.1103/PhysRevLett.37.1251}.

\bibitem[{\citenamefont{Razamat and Tong}(2021)}]{Razamat:2020kyf}
\bibinfo{author}{\bibfnamefont{S.~S.} \bibnamefont{Razamat}} \bibnamefont{and}
  \bibinfo{author}{\bibfnamefont{D.}~\bibnamefont{Tong}},
  \bibinfo{journal}{Phys. Rev. X} \textbf{\bibinfo{volume}{11}},
  \bibinfo{pages}{011063} (\bibinfo{year}{2021}), \eprint{2009.05037}.

\bibitem[{\citenamefont{Hands}(2021)}]{Hands:2021mrg}
\bibinfo{author}{\bibfnamefont{S.}~\bibnamefont{Hands}},
  \bibinfo{journal}{Symmetry} \textbf{\bibinfo{volume}{13}},
  \bibinfo{pages}{1523} (\bibinfo{year}{2021}), \eprint{2105.09646}.

\bibitem[{\citenamefont{Wipf and Lenz}(2022)}]{Wipf:2022hqd}
\bibinfo{author}{\bibfnamefont{A.~W.} \bibnamefont{Wipf}} \bibnamefont{and}
  \bibinfo{author}{\bibfnamefont{J.~J.} \bibnamefont{Lenz}},
  \bibinfo{journal}{Symmetry} \textbf{\bibinfo{volume}{14}},
  \bibinfo{pages}{333} (\bibinfo{year}{2022}), \eprint{2201.01692}.

\bibitem[{\citenamefont{Graf}(1978)}]{Graf:1978kr}
\bibinfo{author}{\bibfnamefont{W.}~\bibnamefont{Graf}}, \bibinfo{journal}{Ann.
  Inst. H. Poincare Phys. Theor.} \textbf{\bibinfo{volume}{29}},
  \bibinfo{pages}{85} (\bibinfo{year}{1978}).

\bibitem[{\citenamefont{Chamseddine}(1990)}]{Chamseddine:1990gk}
\bibinfo{author}{\bibfnamefont{A.~H.} \bibnamefont{Chamseddine}},
  \bibinfo{journal}{Nucl. Phys. B} \textbf{\bibinfo{volume}{346}},
  \bibinfo{pages}{213} (\bibinfo{year}{1990}).

\bibitem[{\citenamefont{Zanelli}(2005)}]{Zanelli:2005sa}
\bibinfo{author}{\bibfnamefont{J.}~\bibnamefont{Zanelli}}, in
  \emph{\bibinfo{booktitle}{{7th Mexican Workshop on Particles and Fields}}}
  (\bibinfo{year}{2005}), \eprint{hep-th/0502193}.

\bibitem[{\citenamefont{Witten}(2007)}]{Witten:2007kt}
\bibinfo{author}{\bibfnamefont{E.}~\bibnamefont{Witten}}
  (\bibinfo{year}{2007}), \eprint{0706.3359}.

\bibitem[{\citenamefont{Callan and Harvey}(1985)}]{Callan:1984sa}
\bibinfo{author}{\bibfnamefont{C.~G.} \bibnamefont{Callan}, \bibfnamefont{Jr.}}
  \bibnamefont{and} \bibinfo{author}{\bibfnamefont{J.~A.}
  \bibnamefont{Harvey}}, \bibinfo{journal}{Nucl. Phys. B}
  \textbf{\bibinfo{volume}{250}}, \bibinfo{pages}{427} (\bibinfo{year}{1985}).

\bibitem[{\citenamefont{Kaplan}(1992)}]{Kaplan:1992bt}
\bibinfo{author}{\bibfnamefont{D.~B.} \bibnamefont{Kaplan}},
  \bibinfo{journal}{Phys. Lett. B} \textbf{\bibinfo{volume}{288}},
  \bibinfo{pages}{342} (\bibinfo{year}{1992}), \eprint{hep-lat/9206013}.

\bibitem[{\citenamefont{Chamseddine}(1989)}]{Chamseddine:1989nu}
\bibinfo{author}{\bibfnamefont{A.~H.} \bibnamefont{Chamseddine}},
  \bibinfo{journal}{Phys. Lett. B} \textbf{\bibinfo{volume}{233}},
  \bibinfo{pages}{291} (\bibinfo{year}{1989}).

\bibitem[{\citenamefont{Witten}(1988)}]{Witten:1988hc}
\bibinfo{author}{\bibfnamefont{E.}~\bibnamefont{Witten}},
  \bibinfo{journal}{Nucl. Phys. B} \textbf{\bibinfo{volume}{311}},
  \bibinfo{pages}{46} (\bibinfo{year}{1988}).

\bibitem[{\citenamefont{Butt et~al.}(2018)\citenamefont{Butt, Catterall, and
  Schaich}}]{Butt:2018nkn}
\bibinfo{author}{\bibfnamefont{N.}~\bibnamefont{Butt}},
  \bibinfo{author}{\bibfnamefont{S.}~\bibnamefont{Catterall}},
  \bibnamefont{and} \bibinfo{author}{\bibfnamefont{D.}~\bibnamefont{Schaich}},
  \bibinfo{journal}{Phys. Rev. D} \textbf{\bibinfo{volume}{98}},
  \bibinfo{pages}{114514} (\bibinfo{year}{2018}), \eprint{1810.06117}.

\bibitem[{\citenamefont{Ojima}(1989)}]{10.1143/PTP.81.512}
\bibinfo{author}{\bibfnamefont{S.}~\bibnamefont{Ojima}},
  \bibinfo{journal}{Progress of Theoretical Physics}
  \textbf{\bibinfo{volume}{81}}, \bibinfo{pages}{512} (\bibinfo{year}{1989}),
  ISSN \bibinfo{issn}{0033-068X}.

\end{thebibliography}

\end{document}